\documentclass[prb,twocolumn,showpacs,preprintnumbers,amsmath,amssymb,superscriptaddress]{revtex4-1}

\usepackage{dcolumn}
\usepackage{bm}
\usepackage{amsmath,graphicx}
\usepackage{color}
\usepackage{bm}
\usepackage{url}

\newcommand{\pa}{\partial}
\newcommand{\ep}{\varepsilon}
\newcommand{\tht}{\theta}

\begin{document}
\title{Scaling law for the onset of the surface wrinkling of multilayer tubes}

\author{Motohiro Sato}
\affiliation{Division of Mechanical and Aerospace Engineering,
Faculty of Engineering, Hokkaido University, Kita 13 Nishi 8, Sapporo 060-8628, Japan}

\author{Kazusa Ishigami}
\affiliation{Division of Socio-Environmental Engineering, Graduate School of Engineering, Hokkaido University, 
N13-W8, Kita-ku, Sapporo, Hokkaido 060-8628, Japan}

\author{Hiroyuki Kato}
\affiliation{Division of Mechanical and Aerospace Engineering,
Faculty of Engineering, Hokkaido University, Kita 13 Nishi 8, Sapporo 060-8628, Japan}

\author{Yoshitaka Umeno}
\affiliation{Institute of Industrial Science, The University of Tokyo, 4-6-1 Komaba, Meguro-ku, Tokyo 153-8505, Japan}

\author{Hiroyuki Shima}
\email{hshima@yamanashi.ac.jp}
\thanks{(Correspondence author)} 
\affiliation{Department of Environmental Sciences, University of Yamanashi, 4-4-37, Takeda, Kofu, Yamanashi 400-8510, Japan}

\date{\today}

\begin{abstract}
Surface wrinkling is an instability mode that is often observed in a wide variety of multilayer tubes under bending deformation. When the degree of applied bending exceeds a critical value, wrinkles appear at the intrados of the tube and release a large amount of in-plane strain energy stored by bending deformation. In the present work, we propose a simple theoretical model for evaluating the critical curvature and critical bending moment for the occurrence of wrinkling in multilayer tubes and apply the model to carbon nanotubes in a case study. Results indicate an inverse proportional relationship between the two critical properties, which holds true regardless of the number of layers and the size of the hollow core.
\end{abstract}

\maketitle

\section{Introduction}

Coaxially stacked tubular structures are omnipresent in nature and technology \cite{Shimizu2005}.
Actual realizations in living matter include
self-assembled microtubules in cells \cite{SeptPRL2010,Brouhard2018},
blood vessels \cite{Silva2016}, and lipid 
tubes \cite{YZhao2005,YZhaoPRE2009,Sekine2012,Ghellab2018},
whose biological functions are attributed to their multilayered tubular structures
\cite{HSShenPLA2010,MTajBBRC2012,YGaoPhysicaE2010}.
In the field of nanotechnology, series of multiwalled nanotubes have been successfully synthesized; 
those made from carbon \cite{ShimaBook2012,ShimaMater2012}, 
boron nitride \cite{CZhiMaterSciEngR2010,MZhengNTN2012}, 
silica \cite{DelclosNanoLett2008}, 
a noble metal \cite{PGaoLangmuir2006,JZhuJNR2013},
WS$_2$ \cite{KaplanAshiriPNAS2006,KalfonCohenJVSTB2011}, 
and TiO$_2$ \cite{JQiuJMaterChem2012} are 
only a few to mention. 
Furthermore, 
coaxial multilayered cylinders have merit in the design of macroscale
composite materials in many engineering fields \cite{Sato2007,SatoSEM2008,Ozbakkaloglu2013}. 

An important benefit of coaxial multilayered tubes,
compared with monolayer tubes, is 
the enhanced mechanical robustness of the original circular cross-section against bending.
When a thin and initially straight monolayer tube is bent gradually,
the tube cross-section first ovalizes \cite{Brazier1927}
and then collapses locally by forming a sharp ridge connecting two kinks 
\cite{Kyriakides1992,Lobkovsky1996}.
These cross-sectional deformations are largely suppressed in multilayered cases.
The suppression is primarily due to the steric effect; {\it i.e.,}
inner tubes push back the inward deflection of outer tubes,
thus preventing the cross-sections from severe deformation under bending.
Above a threshold of bending, however,
the steric effect is insufficient to suppress the deflection
and another type of instability mode is likely to appear,
{\it i.e.,} surface wrinkling.
Surface wrinkling is wavelike distortion along the compressed intrados of the tubes,
through which the compressed intrados is allowed to accommodate the bending-induced compressive strain.
Surface wrinkling is a universal instability mode 
in that it has been observed in a wide variety of deformable materials with high aspect ratios,
such as thick rubber cuboids \cite{Gent1999}, rubber scrolls \cite{Mahadevan2004},
and macroscale monolayer metal tubes \cite{HYang2004}.

To evolve multilayered tube applications,
it is important to identify the conditions under which surface wrinkling occurs;
i.e., we want to identify the number of tubes stacked coaxially and the degree of bending of the tube axis that result in surface wrinkling.
The phenomenon is nonlinear in nature, and an exact
evaluation of the conditions thus requires complicated and expensive numerical simulations,
as was shown in the case of multiwalled carbon nanotubes (MWNTs).
It is widely known that
bending deformation 
strongly affects the physical 
properties of MWNTs \cite{Rochefort1999, Farajian2003,JMa2015,Ouakad2016},
and intense numerical efforts have thus been made to reveal
the surface morphology and energetics of wrinkled MWNTs
\cite{Poncharal1999,Arroyo2003,TChang2006,XYLi2007,XHuang2008,Arroyo2008,Nikiforov2010,CGWang2016,Zare2017}.
In complementing such numerical simulations,
an elastic approximation based on shell theory is an alternative simple and effective approach 
for obtaining the wrinkling nature.

In the present work, we propose a simple theoretical model for estimating the approximate shape 
and bending occurrence threshold of surface wrinkling of multilayer tubes. 
Using the model, we calculate the critical curvature of the tube axis 
and the critical bending moment for the surface wrinkling of MWNTs.
The two critical properties are found to obey a simple rule,
which provides an important guideline for maximum bending deflection.
The applicability of the simple rule
to multilayer tubes with different spatial scales,
even macroscopic multilayer tubes,
is discussed.

\section{Method of analysis}

\subsection{Elastic approximation model}

\begin{figure}[ttt]
\centering
\includegraphics[width=4.5cm]{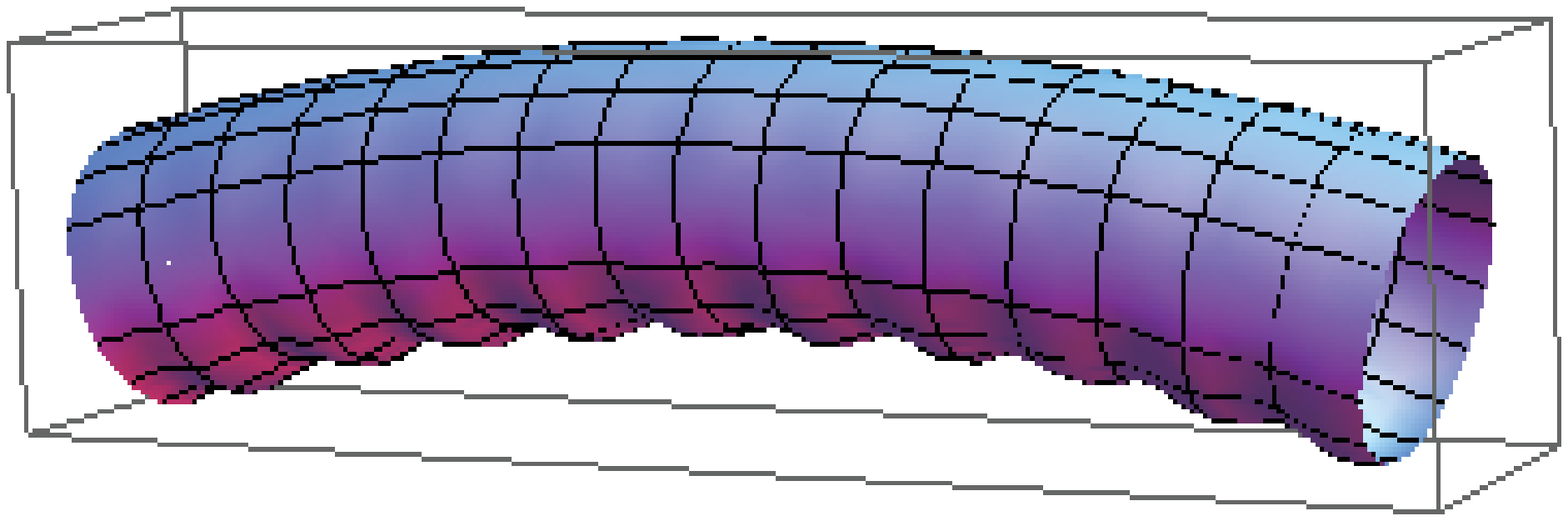}
\includegraphics[width=3.5cm]{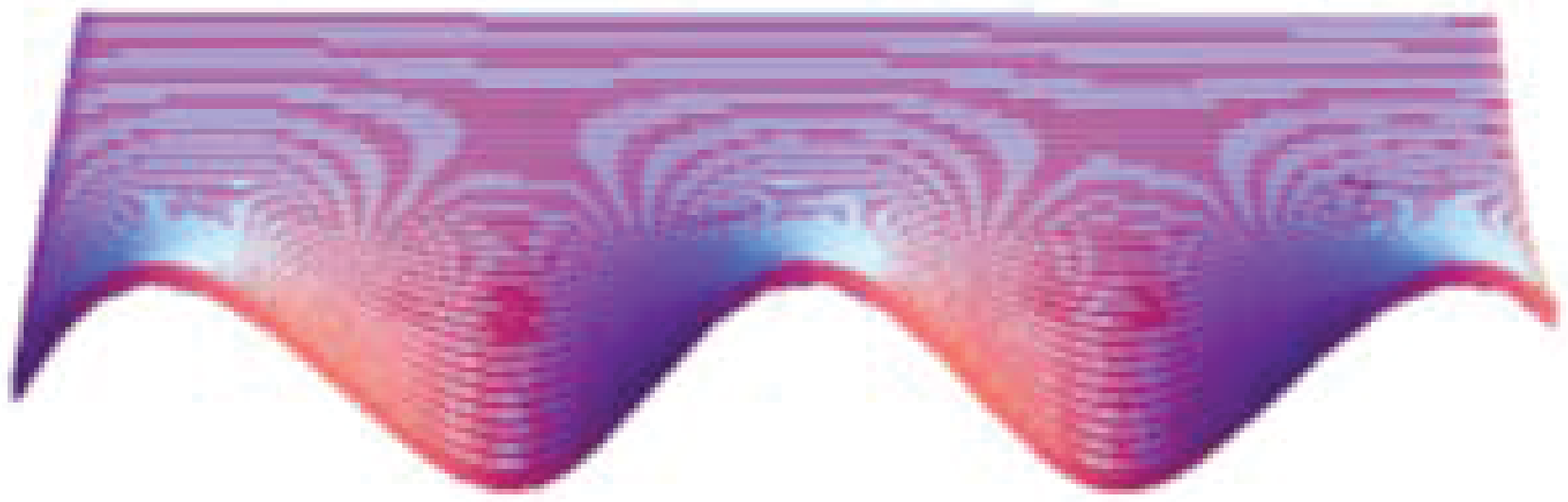}
\caption{Left: Diagonal view of a thick MWNT under bending; only the outmost wall is presented.
A wave-shaped corrugation pattern appears on the compressed side when
a bending moment above a threshold is applied.
Right: Enhanced view of the corrugated intrados.}
\label{fig_01}
\end{figure}

Surface wrinkling has been observed for many multiwalled nanotubes
made of
carbon \cite{Kuzumaki1998,Lourie1998,Poncharal1999,Bower1999,Jackman2014,Jackman2015},
boron nitride \cite{Golberg2007,YHuang2013,Ghassemi2011}, and 
inorganic materials \cite{MSWang2008,Bucholz2012}.
In those nanomaterials, interaction between adjacent monoatomic layers
originates from intermolecular van der Waals (vdW) forces
and is thus weak compared with strong chemical bonding \cite{ShimaBook2012}.
This anisotropy in mechanical stiffness
causes surface wrinkling under bending (see Fig.~\ref{fig_01});
the flexibility of individual layers in response to out-of-plane deflection,
relative to the high rigidity of the individual layers against in-plane deformation, 
allows the release of an appreciable amount of membrane strain energy 
through surface wrinkling
at the expense of a slight increment in vdW energy.

Figure \ref{fig_02} illustrates our analytical model of MWNTs,
comprising many continuum thin elastic tubes with length $L$.
We assume that MWNTs deform with constant curvature under the application of 
pure bending characterized by the bending moment $M$.
$N$ is the number of concentric tubes while 
$r_i$ is the radius of the $i$-th tube defined by
\begin{equation}
r_i = r_1 + (i-1)d.
\label{eq_x006}
\end{equation}
Here, $d$ is the interlayer distance.
The equilibrium spacing between neighboring walls is set to $d$ = 0.3415 nm,
in accordance with the results of previous studies \cite{Girifalco2000}.

\subsection{Strain energy}

\begin{figure}[ttt]
\centering
\includegraphics[width=7.0cm]{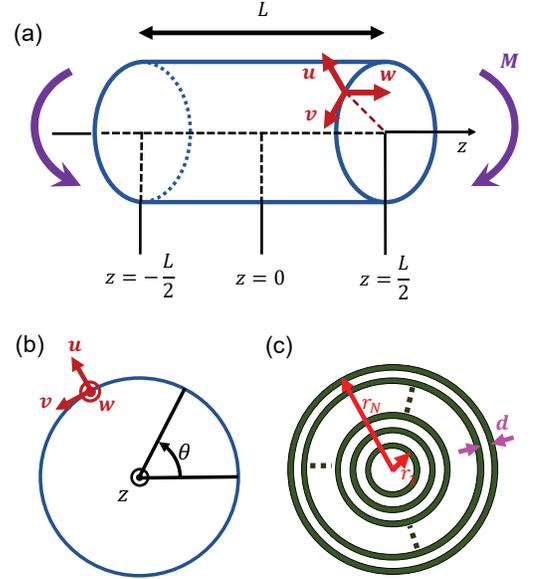}
\caption{(a) Continuum elastic model of an individual tube under bending.
A set of three displacement vectors, $\bm{u}, \bm{v}, \bm{w}$,
and the bending moment $\bm{M}$ are depicted.
(b) Cross-sectional view of the model.
(c) Definitions of the tube radius $r_i$ $(i=1,2,\cdots,N)$
and interwall distance $d$.}
\label{fig_02}
\end{figure}

The cross-sectional shape of an elastic hollow tube under pure bending
is evaluated using thin-shell theory.
The theory states that 
the total strain energy
 $U$ for an $N$-walled tube to deform
is written as the sum of three energy terms;
\color{black}
the general form is given by
\color{black}
\cite{Shima2014}
\begin{eqnarray}
U = \sum_{i=1}^N U_\tht^{(i)} + \sum_{i=1}^N  U_z^{(i)} + {\sum_{i,j}}' U_I^{(ij)}.
\label{eq_01}
\end{eqnarray}
Here, $U_\theta^{(i)}$ is the strain energy 
associated with the circumferential displacement
of a volume element in the $i$-th wall,
$U_z^{(i)}$ is that associated with the axial displacement,
and $U_I$ accounts for the interaction 
between two adjacent walls.
All three energy terms are functions of 
the curvature $\Gamma$ of the deformed tube axis
and the local displacement of the $i$-th wall
represented by a set of mutually perpendicular vectors 
$\bm{u}_i, \bm{v}_i, \bm{w}_i$ (see Fig.~\ref{fig_02}).
Amplitudes of the vectors, denoted by $u_i, v_i, w_i$, 
indicate the displacements of a volume element
in the radial, circumferential, and axial directions, respectively.
\color{black}
Specifically when considering the surface wrinkling of $N$-walled sysmtems 
with $N>>1$, we are allowed to set $v_i\equiv 0$ for every $i$ as explained later.
But for the time being, we will proceed with general theory 
without making this assumption.
\color{black}

The circumferential strain energy $U_\tht^{(i)}$ of the $i$-th wall
is explicitly given by
\begin{equation}
U_\tht^{(i)} = 
\frac{r_i}{2} \int_0^L dz \int_0^{2\pi} d\tht
\left( \frac{C}{1-\nu^2} \ep_{\tht}^2 + D \kappa_{\tht}^2 \right),
\label{eq_x001}
\end{equation}
where
\begin{equation}
\ep_{\tht} = \frac{u_i + \pa_\tht v_i}{r_i} 
+
\frac{\left( \pa_\tht u_i - v_i \right)^2}{2 r_i^2},
\;\;
\kappa_{\tht} = - \frac{\pa_\tht^2 u_i - \pa_\tht v_i}{r_i^2}.
\label{eq_x002}
\end{equation}
The symbol $\pa_\tht$ indicates the partial derivative with respect to $\tht$.
The constant $C$ denotes the in-plane stiffness,
$D$ the flexural rigidity, 
and $\nu$ the Poisson ratio of each wall.
In actual calculations, we substitute $C = 345$ nN/nm,
$D = 0.238$ nN$\cdot$ nm, and $\nu =0.149$
by following the previous work \cite{Kudin2001} based on density functional theory.

In a similar manner, the axial strain energy $U_z^{(i)}$ is given by
\begin{equation}
U_z^{(i)} = 
\frac{r_i}{2} \int_0^L dz \int_0^{2\pi} d\tht
\left( \frac{C}{1-\nu^2} \ep_z^2 + D \kappa_z^2 \right),
\label{eq_x003}
\end{equation}
with the definitions
\begin{equation}
\ep_z = \pa_z w_i + \frac{\left( \pa_z u_i \right)^2}{2} + \frac{\left( \pa_z v_i \right)^2}{2},
\quad \kappa_z = \pa_z^2 u_i.
\label{eq_x004}
\end{equation}
The interaction energy is defined by
\begin{equation}
U_I^{(ij)} = 
\frac{\chi_{ij} (r_i+r_j)}{4} \int_0^L dz \int_0^{2\pi} d\tht_i
(u_i-u_j)^2 ,
\label{eq_x005}
\end{equation}
where $\chi_{ij}$ is the effective spring constant per surface area, 
which serves as a measure of the vdW interaction strength.
The explicit function form of $\chi_{ij}$ and its derivation
were detailed in Ref.~\cite{ShimaCMS2012}.

It is noteworthy that the values of $C$ and $D$ depend on the tube radius, in principle.
Nevertheless, the dependencies become negligible when the tube radius exceeds 0.5 nm, 
above which the elastic constants of carbon nanotubes converge to those of 
a planar graphene sheet \cite{Kudin2001}. 
Against this background, we consider only the nanotubes 
whose radii are larger than 0.5 nm, 
which allows us to fix the values of $C$ and $D$ as noted above.

\subsection{Surface wrinkling mode analysis}

To identify the bending threshold for the occurrence of surface wrinkling,
we decompose the displacement of a volume element in the $i$-th tube as 
\begin{eqnarray}
u_i(z,\tht) &=& u_i^*(z,\tht) + \delta u_i(z,\tht), \nonumber \\
v_i(z,\tht) &=& v_i^*(z,\tht) + \delta v_i(z,\tht), \nonumber \\
w_i(z,\tht) &=& w_i^*(z,\tht) + \delta w_i(z,\tht). \label{eq_decomp}
\end{eqnarray}
The symbols marked by an asterisk are displacements in the stable mode,
during which wrinkling has not yet happened such that
the cross-section remains uniform along the tube axis.
Those marked by $\delta$ are infinitesimal displacements 
just after the occurrence of the surface wrinkling mode,
which are to be observed
immediately after the degree of bending exceeds
a critical threshold.

We make a few assumptions for 
the displacement components to simplify the model.
First, we recall that,
when considering many-walled nanotubes with $N\gg 1$, 
the ovalization effect observed in the stable mode 
can be small and negligible \cite{Shima2014}. 
We thus set $u_i^*=0$ and $v_i^*=0$ for simplicity. 
Second, we assume that the axial displacement in the stable mode is given by
\begin{equation}
w_i^* = r_i \Gamma \left( z-\frac{L}{2} \right) \sin\theta,
\end{equation}
taking into account 
the elongation and shrinkage of the extrados (at $\theta=\pi/2$) and intrados ($\theta=-\pi/2$),
respectively, as a result of pure bending.
Third, in the wrinkling mode analysis, 
we ignore the contributions of $\delta v_i$ and $\delta w_i$ to the strain energy,
because they have minor effects on the surface morphology of thick MWNTs \cite{Shima2014}.
We thus focus only on the pronounced contribution of $\delta u_i$.
Finally, we expand $\delta u_i$ using a Fourier cosine series with respect to $\tht$
up to second order as
\begin{equation}
\delta u_i = 
\left[
a_i + b_i \cos \left( \theta-\frac{\pi}{2} \right) + c_i \cos 2\theta
\right] 
\sin \left( \frac{n\pi}{L} z \right).
\label{eq_085}
\end{equation}
In the second term in the square brackets, 
the phase shift of $-\pi/2$ is artificially added 
to achieve the situation that the amplitude of $\delta u_i$
at the compressed intrados is maximized.
The integer $n$ defined in Eq.~(\ref{eq_085}) indicates the number of waves
on the compressed intrados.

Substituting the function forms of $u_i, v_i, w_i$ described above into Eq.~(\ref{eq_01}) and 
rearranging the equation with respect to the order of three Fourier coefficients, $a_i, b_i, c_i$,
we obtain the expression
\begin{equation}
U = U^* + \delta U \left( a_i, b_i, c_i \right) + \delta^2 U \left( a_i^2, b_i^2, c_i^2 \right).
\end{equation}
To determine the critical buckling point,
we impose the stability condition that the differentiation of
the second variation in strain energy, $\delta^2 U$,
with respect to the Fourier expansion coefficients should be zero:
\begin{equation}
\frac{\partial (\delta^2 U)}{\partial a_i} = 0,\;\;
\frac{\partial (\delta^2 U)}{\partial b_i} = 0,\;\;
\frac{\partial (\delta^2 U)}{\partial c_i} = 0.
\label{eq_105}
\end{equation}
Solving Eqs.~(\ref{eq_105}) under the conditions of $a_i^2+b_i^2+c_i^2\ne 0$ for all $i$
gives a series of curvature $\Gamma$ as a function of $n$,
each of which may cause surface wrinkling with wavenumber $n$.
Among them, the minimum value of $\Gamma$
corresponds to the critical curvature $\Gamma_{\rm cr}$,
which is to be observed in experiments.
The critical bending moment, $M_{\rm cr}$, is given by
\begin{equation}
M_{\rm cr} = \pi C \Gamma_{\rm cr} \sum_{i=1}^N (r_i)^3.
\end{equation}

\begin{figure}[ttt]
\centering
\includegraphics[width=4.2cm]{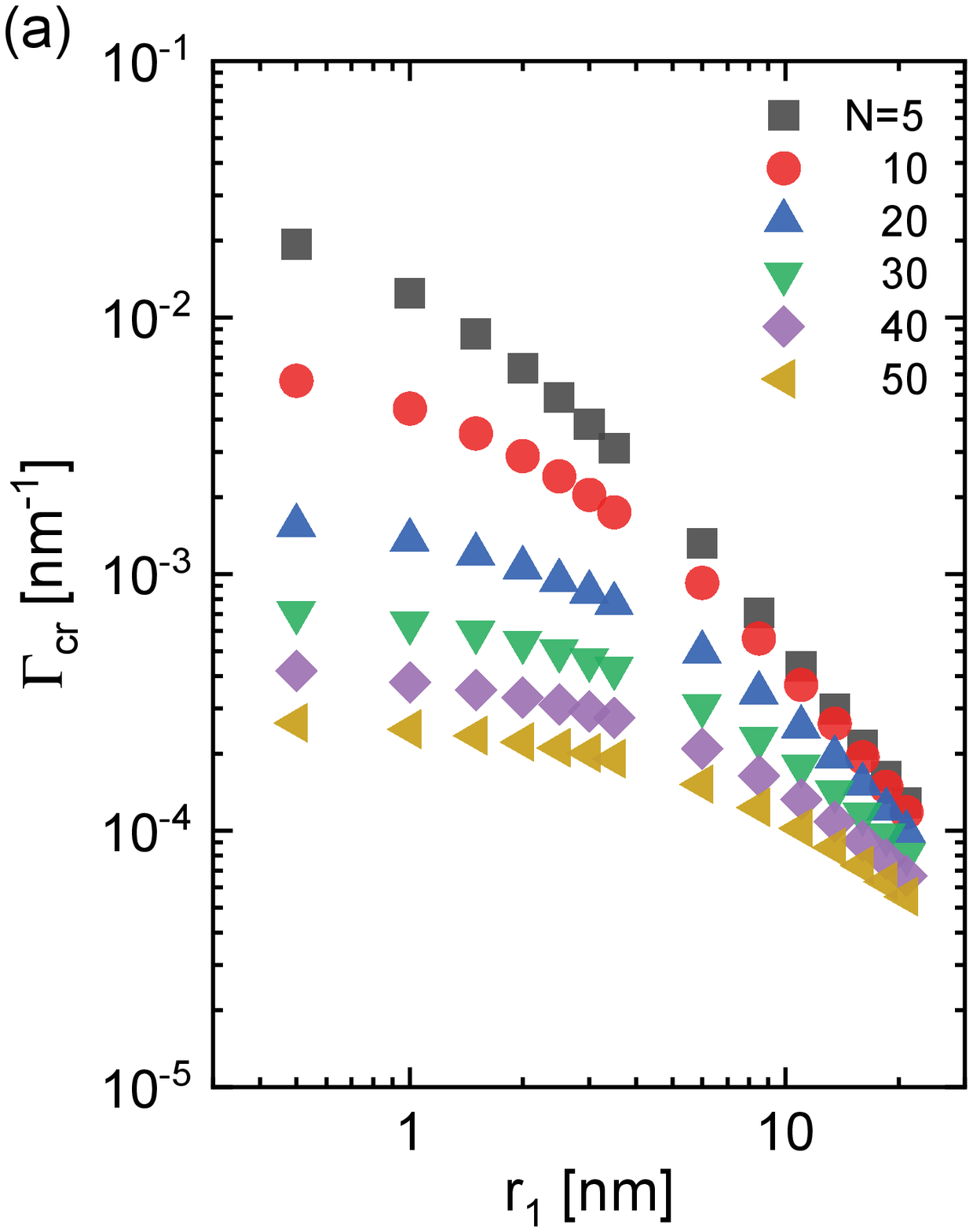}
\includegraphics[width=4.2cm]{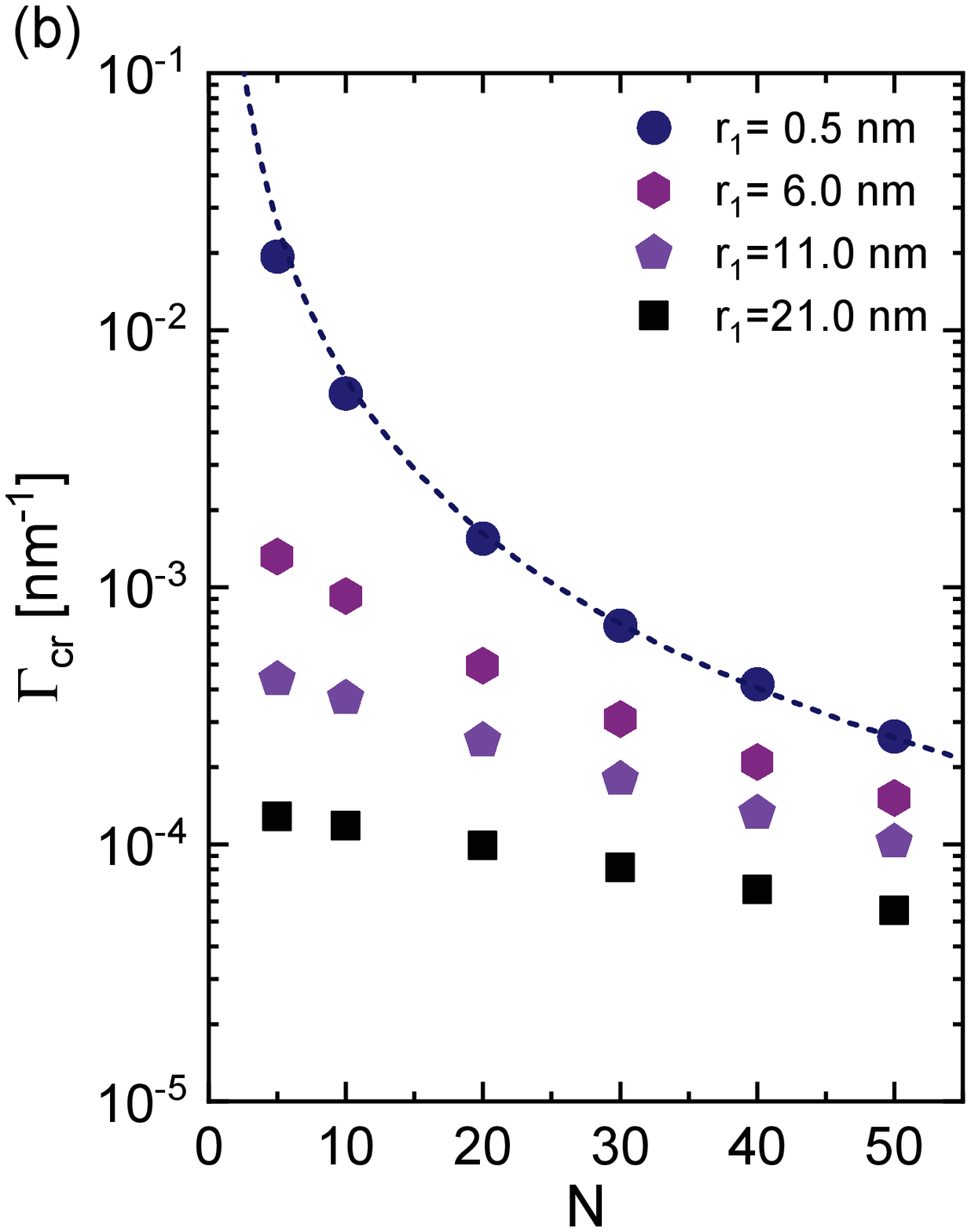}
\caption{(a) Critical bending curvature $\Gamma_{\rm cr}$
as a function of (a) the innermost wall radius $r_1$ and
(b) the number of walls $N$.
The dotted curve obeys a power law of $\Gamma_{\rm cr} \propto N^{-2}$.}
\label{fig_03}
\end{figure}

\begin{figure}[ttt]
\centering
\includegraphics[width=4.2cm]{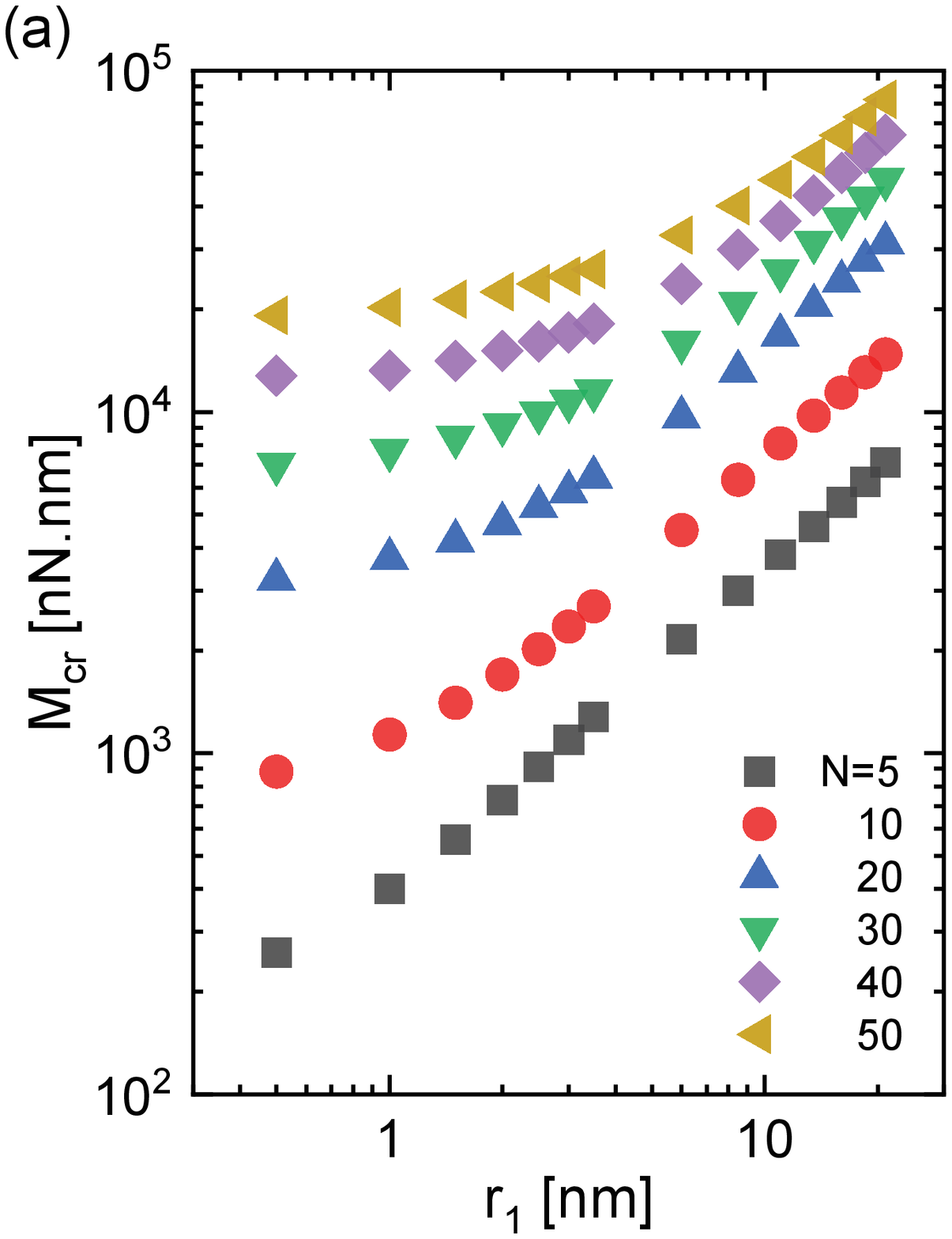}
\includegraphics[width=4.2cm]{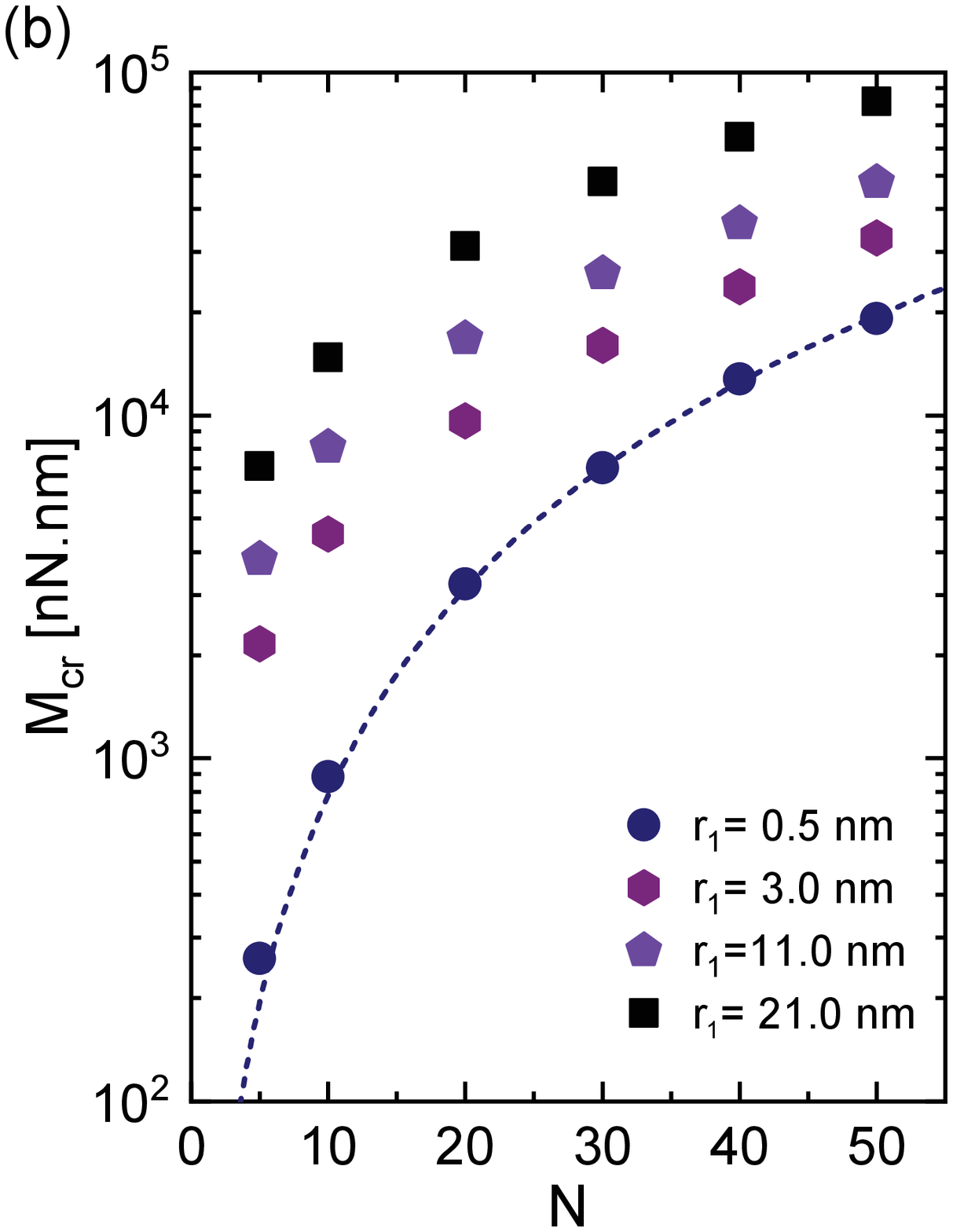}
\caption{(a) Critical bending moment $M_{\rm cr}$
as a function of (a) the innermost wall radius $r_1$ and
(b) the number of walls $N$.
The dotted curve obeys a power law of $M_{\rm cr} \propto N^2$.}
\label{fig_04}
\end{figure}

\section{Results}

\subsection{Critical curvature}

Figure \ref{fig_03}(a) shows the critical curvature $\Gamma_{\rm cr}$ of $N$-walled nanotubes
as a function of $r_1$.
In actual calculation, 
the innermost tube radius $r_1$ is increased from 0.5 to 21.0 nm
and the number of constituent walls $N$ is varied between 5 and 50;
the tube length $L$ is fixed at 102.0 nm.
In Fig.~\ref{fig_03}(a),
a monotonic decrease in $\Gamma_{\rm cr}$ with increasing $r_1$ is observed for each value of $N$,
indicating that 
an increase in the outmost tube radius $r_N$ 
reduces the degree of bending that suffices for surface wrinkling to happen.
For instance, the system with $r_1=0.5$ nm and $N=5$ ({\it i.e.,} $r_N\simeq 3.4$ nm) 
shows wrinkling in the event of pure bending
defined by $\Gamma\simeq 2\times 10^{-2}$ nm$^{-1}$,
which corresponds to the curvature radius of the tube axis being on the order of 100 nm,
comparable to the tube length.
Meanwhile, the critical curvature radius
for the system with $r_1=21.0$ nm and $N=5$ ({\it i.e.,} $r_N\simeq 16$ nm) 
is estimated to be $\sim 10$ $\mu$m,
which is 100 times that in the previous case.
Similar trends are found in the results for each value of $N$.

Figure \ref{fig_03}(b) illustrates how $\Gamma_{\rm cr}$ depends on the number of walls $N$
when $r_1$ is fixed.
It follows that for a smaller $r_1$,
the data are well fitted by a negative power law with an exponent of $-2$.
Even for a larger $r_1$,
this power law applies if the net thickness of an $N$-walled tube, defined by $(N-1)d$ 
or equivalently $r_N-r_1$,
is larger than the innermost tube radius $r_1$.

\subsection{Critical bending moment}

Figure \ref{fig_04}(a) presents the $r_1$ dependence of the critical bending moment $M_{\rm cr}$ 
of $N$-walled nanotubes.
For each value of $N$, $M_{\rm cr}$ is found to increase monotonically with increasing $r_1$.
The monotonic increase indicates 
the necessity of applying a larger bending moment
for wrinkles to emerge when bending a thicker MWNT.
In other words, it is a consequence of the stiffening effect
against bending achieved by increasing the radius of the outermost tube.

Figure \ref{fig_04}(b) shows rapid increases in $M_{\rm cr}$ with $N$,
where the data for the small $r_1$ collapse onto a power law with an exponent of 2.
Again, this power law holds true for the system satisfying $(N-1)d > r_1$, in which the entire system can be 
considered as densely stacked multilayered tubes with a very thin cavity.

\begin{figure}[ttt]
\centering
\includegraphics[width=7.4cm]{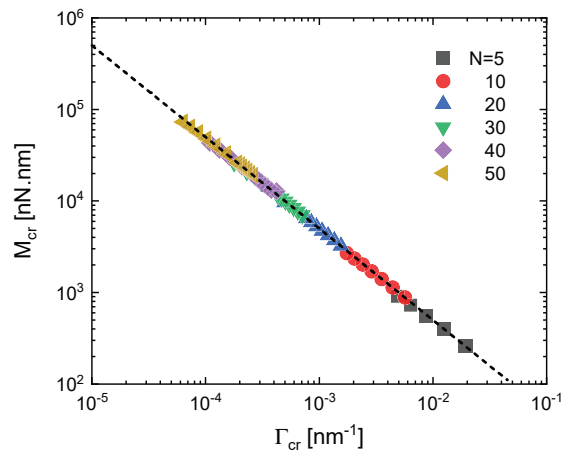}
\caption{Log--log plot of the $M_{\rm cr}$--$\Gamma_{\rm cr}$ curve. 
Data collapse onto a unified power law of $M_{\rm cr} \propto \Gamma_{\rm cr}^{-1}$.}
\label{fig_05}
\end{figure}

\subsection{Scaling law}

We demonstrate that two characteristic quantities of surface wrinkling, 
$M_{\rm cr}$ and $\Gamma_{\rm cr}$,
obey a simple scaling law under a certain condition.
Figure \ref{fig_05} shows a log--log plot of $M_{\rm cr}$ and $\Gamma_{\rm cr}$
derived from systems satisfying the densely stacked criterion that $(N-1)d > r_1$.
The graph clearly shows that all data collapse onto 
a simple relation given by 
\begin{equation}
M_{\rm cr}\Gamma_{\rm cr} = k,
\label{eq_135}
\end{equation}
with the constant $k=5.0$ nN,
as long as the criterion is satisfied.
Among the data points calculated from the systems having the same number of tubes $N$, 
those located in the upper left (or lower right) are obtained from MWNTs having a small (large) $r_1$.

\section{Discussion}

\color{black}
\subsection{Scale invariance of the constant $k$}
\color{black}

The following discussion derives the scaling law of Eq.~(\ref{eq_135})
using an approximation theory and proves that
the constant $k$ given in Eq.~(\ref{eq_135}) is scale invariant
and thus dependent on neither $N$ nor $r_1$.

We first recall that 
the buckling behavior of {\it axially compressed} circular cylindrical shells
has been a subject discussed for many decades in the field of structural mechanics.
When a cylindrical shell with a long slender shape
is compressed axially,
it buckles sideways as a whole.
The situation is different for a shorter cylindrical shell; {\it i.e.},
when a cylindrical shell with a radius $R$ and a medium length $L\simeq R$
is compressed axially,
the shell buckles locally
while the cylindrical axis remains straight \cite{Feliciano2011}.
The critical strain $\ep_{\rm cr}$ 
for such local buckling in the Lorenz limit \cite{Lorenz1911} reads
\begin{equation}
\ep_{\rm cr} = \frac{2}{R} \sqrt{\frac{D}{C}},
\label{eq_140}
\end{equation}
where $C$ and $D$ are respectively the in-plane stiffness and flexural rigidity
of the shell.

We now hypothesize that Eq.~(\ref{eq_140}) applies to 
local deformation ({\it i.e.,} surface wrinkling)
of the compressed side of the outermost wall of a MWNT under pure {\it bending}.
The critical strain for surface wrinkling at the compressed intrados
may then be described by
\begin{equation}
\ep_{\rm cr}(N,r_1) = \frac{2}{r_N(N,r_1)} \sqrt{\frac{D}{C}},
\end{equation}
where the presence or absence of the dependence on the variables $N$ and $r_1$
is explicitly shown.
Considering the geometric meaning of $\Gamma_{\rm cr}$, 
$\ep_{\rm cr}(N,r_1)$ should satisfy the relation 
\begin{equation}
\ep_{\rm cr}(N,r_1) = r_N(N,r_1) \Gamma_{\rm cr}(N,r_1).
\end{equation}
Eliminating $\ep_{\rm cr}$ from the two equations above, we obtain
\begin{equation}
\Gamma_{\rm cr}(N,r_1) = \frac{2}{\left[ r_N(N,r_1) \right]^2} \sqrt{\frac{D}{C}},
\end{equation}
which implies
\begin{equation}
\Gamma_{\rm cr}(N, r_1) \left[ r_N(N, r_1) \right]^2 = 2\sqrt{\frac{D}{C}} = {\rm const}.
\label{eq_138}
\end{equation}
It is emphasized that 
the constant on the right hand side, $2\sqrt{C/D}$,
is independent of both $N$ and $r_1$.

As to the critical bending moment, it follows from definition that
\begin{equation}
M_{\rm cr}(N, r_1) = \pi C \Gamma_{\rm cr}(N, r_1) \sum_{i=1}^N \left[ r_i(r_1) \right]^3.
\label{eq_145}
\end{equation}
When $(r_N-r_1)/r_1 > 1$ and $N\gg 1$,
the sum of the cubes of $r_i$ involved on the right-hand side of Eq.~(\ref{eq_145})
can be approximately expressed by
\begin{equation}
\sum_{i=1}^N \left[ r_i(r_1) \right]^3 \simeq \frac{\left[ r_N(N, r_1) \right]^4}{4d},
\label{eq_155}
\end{equation}
as can be derived from
\begin{equation}
\sum_{i=1}^N \left( \frac{r_i}{r_N} \right)^3 
= \sum_{i=1}^N \left[ \left(1-\frac{Nd}{r_N} \right) + \frac{d}{r_N}i \right]^3
\simeq \frac{N}{4},
\end{equation}
where $r_i = r_N - (N-i)d$ has been imposed on the left-hand side.

From Eqs.~(\ref{eq_138})--(\ref{eq_155}), we have
\begin{equation}
M_{\rm cr}(N, r_1) \left[ r_N(N,r_1) \right]^{-2} = \frac{\pi \sqrt{CD}}{2 d} = {\rm const}
\end{equation}
and
\begin{equation}
M_{\rm cr}(N, r_1) \Gamma_{\rm cr}(N, r_1) = \frac{\pi D}{d}.
\label{eq_234}
\end{equation}
We eventually conclude that the constant $k$ 
given in Eq.~(\ref{eq_135}) is approximately equal to $\pi D/d$,
which involves neither $N$ nor $r_1$ as expected.

An intriguing physical consequence deduced from Eq.~(\ref{eq_234}) is that 
the product of $M_{\rm cr}$ and $\Gamma_{\rm cr}$
is determined only by $D/d$; {\it i.e.},
the characteristic force that is required to transform 
a thin elastic plate with flexural rigidity $D$
into a cylinder with radius of curvature $d$.
This result implies that 
if elastic membranes having flexural rigidity $D$ are coaxially stacked 
with the same membrane spacing $d$, 
the resulting multilayer tube follows the scaling law of Eq.~(\ref{eq_135}),
regardless of the number of membranes stacked.

\color{black}
\subsection{Occurrence conditions for the scaling law}

We have demonstrated that, for the scaling law of Eq.~(\ref{eq_135}) to occur, many hollow tubes must be densely stacked around a common tubular axis; this condition is mathematically expressed by $(N-1)d > r_1$. Under the densely stacked condition, the presence of many inner tubes suppresses inward deformation of outer tubes so that the original circular cross-section remains kept during bending until the surface wrinkling takes place. 
Particularly in the case of MWNTs' model, the effective spring constant, $\chi_{ij}$ in Eq.~(\ref{eq_x005}), needs to be larger than a threshold to prevent the inward penetration of the outer tube. 
But we should note that the effective spring constant is not a unique way to represent the interlayer coupling; any kind of interaction will result in the scaling law as long as it suppresses the cross-sectional deformation under the condition of $(N-1)d > r_1$. In multilayered rubber tubes \cite{Mahadevan2004}, for instance, adjacent layers do not exert forces on each other when they are separated; the steric exclusion does work only when they are in contact with each other. Still in the latter case, what is necessary for the scaling law realization will be the densely stacked condition, similar to the case of MWNTs.

Another necessary condition for the scaling law to occur
is that the system should be long and slender enough so that 
the shell theory can apply to the analysis of buckling behaviors under bending.
It was argued that, given an elastic hollow cylinder with length $L$ and diameter $2r$,
its slenderness should be quantified
not by the aspect ratio $L/(2r)$ but the dimensionless parameter: \cite{Shima2016,JWang2020}
\begin{equation}
\Omega = \sqrt{\frac{L^2 h}{r^3}},
\end{equation}
with $h$ being the wall thickness of the hollow cylinder.
When $\Omega$ is larger than 0.5 or a bit more, the cylinder can be considered slender enough so that 
the shell theory applies \cite{JWang2020}; 
as a consequence, the surface wrinkling occurs in the $N$-walled systems with $N\gg 1$.
Given this criterion, most of the MWNTs we have dealt with 
can be classified to slender systems,
provided that the individual wall thickness be $h\sim 0.34$ nm.
Nevertheless, the abovementioned is only a guideline and not a strict criteria for the slenderness of MWNTs,
because the wall is made out of a monoatomic-thick layer and thus the notion of a wall thickness be elusive.
\color{black}

\color{black}
\subsection{Possible means to verify the scaling law}


The remaining important question is to verify that the theoretically predicted scaling law holds in real MWNTs. A possible way for verifying it by experiments is to directly observe the bending-induced deformation process using an electron microscope. In fact, many in-situ nanomechanical experiments have been conducted so far on the bending process of MWNTs \cite{Kuzumaki1998,Lourie1998,Poncharal1999,Bower1999,Jackman2014,Jackman2015}; a cantilevered nanotube was deflected inside an electron microscope, by which the tube geometry under bending and the loading imposed were accurately measured. Using the technique, it will be possible to measure the critical bending moment and critical curvature of a given MWNT, from which we can verify the applicability of the scaling law we have derived. The same applies to nanoscale multiwalled tubes other than those made of carbon, such as boron nitride nanotubes \cite{Ghassemi2011}.

Large-scale numerical simulation is also a powerful tool for verifying the scaling law in multiwalled nanotube systems. For example, it has been demonstrated that MD calculations can be used to simulate the surface wrinkling of 10-walled nanotubes composed of hundreds of thousands of carbon atoms \cite{XLi2007}. The coarse-grained method can also be used to analyze MWNTs containing by far more atoms \cite{Arroyo2003,Arroyo2008}. In addition, the finite element method was proven to be useful for reproduction of the surface wrinkling of MWNTs obtained by electron micrograph images \cite{Pantano2004}. We believe that these computational approaches would enable us to verify the scaling law for various kinds of multiwalled nanotubes.

\color{black}

We conjecture that the scaling law of Eq.~(\ref{eq_135}) holds true 
even for macroscopic multilayer tubular systems,
considering the generality of our theoretical approach.
\color{black}
Indeed, the surface wrinkling was observed when bending a rubber multilayer tube 
that is made by rolling up a thin palm-sized rubber sheet \cite{Mahadevan2004}.
Similar wrinkling mode will be observable in Kapton tubes \cite{QXJi2017} and gel cylinders \cite{Ghatak2007}
if they are stacked coaxially to realize multilayered tubular systems.
In such the macroscopic tubular systems, the measurement of the critical bending moment and 
the critical curvature is easier than in the nanotube systems.
They thus allow to test whether the scaling law applies 
not only to nanoscopic but also macroscopic systems.
Theoretical analysis on the surface wrinkling of a long rubber block \cite{Padukka2012}
may also be a clue to address the problem.
\color{black}

Furthermore, the wrinkling mode is expected 
when the hollow cavity of a thin-walled monolayer tube is filled with 
an elastic medium with high bulk modulus ({\it e.g.}, a liquid or powder),
because almost no cross-sectional ovalization occurs 
even when a bending stress is applied. 
The same is true for thick monolayer tubes 
where the radius of the hollow cavity is sufficiently smaller than 
the tube radius.
It is therefore interesting to scrutinize the generality 
of the scaling law for the surface wrinkling of various multilayer tubes
that cover a wide range of spatial scales;
the results may give us a common rule
for the mechanics of multilayer tubes under bending loading.

\section{Conclusion}

We constructed a simple theoretical model 
to describe the surface wrinkle phenomenon that occurs in MWNTs 
with pure bending. 
Using this model, we derived the inverse proportional law 
of the critical curvature $\Gamma_{\rm cr}$ and critical bending moment $M_{\rm cr}$ 
and gained the theoretical insight that the proportionality factor has 
a scale-free property independent of the outer ($r_N$) and inner ($r_1$) tube radii
and the number of graphene walls $N$.
It is expected that our theoretical model is basically applicable to a broad class of multilayer tube structures,
and not limited to MWNTs, by imposing appropriate material parameters and interlayer energy expressions
specific to the systems to be considered.
We hope that the results presented will contribute to the development of 
applied technologies for multilayer tubes.

\section*{Acknowledgments}
This work was supported by JSPS KAKENHI Grant Numbers 
18H03818, 18K18801, 19H02020, 19H05359, and 19K03766.

\bibliographystyle{apsrev4-1}
\bibliography{MWNT_Wrinkle_arXiv}

\end{document}